\documentclass[aps,prd,twocolumn,showpacs,showkeys]{revtex4-1}

\usepackage{graphicx}
\usepackage{dcolumn}
\usepackage{bm}
\usepackage{filecontents}
\usepackage{natbib}
\usepackage{caption}
\usepackage{subcaption}
\usepackage{braket}
\usepackage{hyperref}
\usepackage{amssymb,amsmath}
\usepackage{hyperref}

\begin{document}


\title{The Leggett--Garg $K_3$ quantity discriminates  Dirac and Majorana  neutrinos}

\author{Monika Richter}%
 \email{monika.richter@smcebi.edu.pl}
\affiliation{%
Institute of Physics, University of Silesia in Katowice, Katowice, Poland 
}%
\author{Bartosz Dziewit}
 \altaffiliation{bartosz.dziewit@us.edu.pl}
\affiliation{%
Institute of Physics, University of Silesia in Katowice, Katowice, Poland 
}%

\author{Jerzy Dajka}
 \altaffiliation{jerzy.dajka@us.edu.pl}

\affiliation{%
Institute of Physics, University of Silesia in Katowice, Katowice, Poland and\\
Silesian Center for Education and Interdisciplinary Research,
University of Silesia in Katowice, Chorz\'ow, Poland
}%


\begin{abstract}
The   $K_3$ quantity, introduced in a context of the Leggett--Garg  inequality violation, is studied for the neutrino oscillations in matter with phenomenologically modelled dissipative environment. It is shown that the $K_{3}$ function acquires different values depending on whether neutrino is Dirac or Majorana particle, provided that there  is a dissipative interaction between matter and neutrinos.  The difference occurs for various matter densities and can serve as a potential quantifier verifying  the neutrino nature. Moreover, working within phenomenological model one can suggest the values of the matter density and dissipation for which the difference is the most visible. There exist also special conditions in which the violation of the Leggett--Garg inequality, to a different extent for both kinds of neutrino, is observed.  
\end{abstract}

\maketitle


\section{\label{sec:level1}Introduction}

For a long time there were two central controversies   in neutrino physics.  The first was the (already solved)
problem of neutrino masses. Neutrinos  are  massive~\cite{white,data}.   
The  second (still alive)  concerns  neutrino  nature  which can be either Dirac, with particles and antiparticles being different, or  Majorana, with  particles  and  antiparticles  being  the same as it is  in the case of  photons. Measuring properties of such  fragile and weakly interacting particles as neutrinos  is a challenge for experimentalists. One can invoke here a short but embarrassing history of faster-than-light neutrino anomaly. That is why the proposal of a potentially helpful quantifier  credibly indicating desired neutrino's property  is of interest for both theoretical and experimental physicists. There are proposals of solving the Dirac--Majorana neutrino controversy~\cite{propp}. In particular, interferometric pattern~\cite{expp1,gutier} or geometric phases of various type~\cite{bea1,Dajka} can serve as a quantifiers of Majorana and Dirac neutrino's difference.  

In 1964 Bell proposed a way to experimentally 
exclude deterministic local (hidden variable theories) as a possible
interpretation of quantum mechanics~\cite{epr}. His celebrated inequality has been later verified in the experiments performed on the pairs of entangled photons (\cite{clauser},\cite{aspect}) which provided then quite convincing evidence against the hidden local-variable theories. Quantum entanglement has also been studied also in a context of oscillating neutrino systems~\cite{blas}.
%
%
In 1984 another test of the quantum mechanics principles has been developed \cite{podstawa}: while the Bell inequality questioned the correlations between spatially separated systems, the Leggett-Garg inequalities (LGI) put the limits on the time correlations. They were derived basing on the assumptions of \emph{macroscopic realism}  (MR) and \emph{non-invasive measurability} (NIM), the first of which presupposes that the system can be only in one of the available set of states, whereas the second one claims that it is possible to perform a measurement without disturbing  a system \cite{review,fil}. 
Just as the Bell inequalities, Leggett-Garg inequalities have been reported to be violated in the variety of systems (see for instance \cite{experiment2},\cite{experiment1}). Especially interesting in terms of a study of the Leggett-Garg inequalities seem to be the neutrinos oscillations, for which the coherence length (the distance beyond  which, the interference of different massive neutrinos is not observable \cite{Giunti}) is much bigger than for the other systems. Such research has been already made both from the theoretical (\cite{leggettneutrinos}, \cite{leggettneutrinos2}) and experimental points of view -- cf. Ref.~\cite{kaiser} reporting violation of the LGI in neutrino systems based upon the MINOS experiment's data. 

Our aim in this work is essentially more modest: instead of searching for violations of neutrino's macro--realism our primary aim is to present a quantifier which, being (at least in principle) measurable,  can discriminate between Dirac and Majorana neutrinos. 

In this work, complementary to recent studies of neutrinos conducted from the perspective of quantum information~\cite{bea2}, we propose such a quantifier. We show that the $K_3$ correlator, used in the  simplest class of LGIs calculated for neutrinos oscillating in matter in a presence of decohering interaction, has different value for both kinds of neutrinos.
A possible mechanism causing decoherence in neutrino oscillations originate from quantum gravity effects as suggested in Refs.~\cite{gr1,gr2,gr3}. Instead of complex modelling in this work, following~\cite{ben1}, we adopt Kossakowski--Lindblad--Gorini--type phenonological equations. This type of describing open systems is particularly useful in quantum optics~\cite{breu} but nevertheless its is very general as it can be applied to any completely positive dynamics of quantum systems~\cite{alicki}. Although there certain problems of open quantum dynamics where one can directly relate parameters of of the Lindblad equation to the microscopic system--plus--environment description of a considered model~\cite{alicki}, in a case of neutrino oscillation our modelling is purely phenonemological and the effect of decoherence is encoded in a Kossakowski matrix describing non--unitary character of  quantum dynamics. 

Furthermore, we show that certain form of Kossakowski matrix describing decoherence is a {\it sine qua non} condition for usefulness of the $K_3$ function in determining the neutrino's nature. 
We also theoretically investigate  the behaviour of the LGI violation for  oscillating neutrinos treated as an open quantum system  undergoing the dissipation and decoherence while interacting with an environment. 

The paper is organized as follows: first (in Sec. II) for  self--consistency of the presentation we review dissipative dynamics of the neutrino oscillations following~\cite{Benatti,Dajka}, then in Sec. III, we introduce the notion of the Leggett--Garg inequalities~\cite{review} simultaneously providing the conditions for which the $K_3$ correlator allows to discriminate between Dirac and Majorana neutrinos. Finally, we summarize our work. 

\section{Dissipative neutrino oscillations in matter}

As the neutrino of the given flavour $\alpha=\{e,\mu,\tau\}$ is produced at some point, it propagates further as the coherence mixture of 3 massive states. Mathematically, it means that neutrino flavour states $\{\ket{\nu_{e}}, \ket{\nu_{\mu}}, \ket{\nu_{\tau}}\}$ can be considered the linear combinations of the massive states $\{\ket{1}, \ket{2}, \ket{3}\}$. The correspondence between them  is expressed via unitary transformation  sometimes referred to as the lepton mixing matrix $U$ \cite{Giunti}:
\begin{equation}
\label{mixingmatrix}
\left(\begin{array}{c} \ket{\nu_{e}}\\  \ket{\nu_{\mu}}\\ \ket{\nu_{\tau}}\end{array}\right)=U  \left(\begin{array}{c} \ket{1}\\  \ket{2}\\ \ket{3}\end{array}\right).
\end{equation}

In this paper, we limit ourselves only to the two neutrino oscillations case (between $e$ and $\mu$). Since the third neutrino effectively decouples it seems to be a quite reasonable assumption \cite{Smirnov}. Then $U$ matrix can be parametrized by one mixing angle $\theta_{12}=\theta$ and in the most general case  by one CP violating phase $\phi$:

\begin{equation}
\label{Eq1}
\left(\begin{array}{c}\ket{\nu_{e}}\\\ket{\nu_{\mu}}\end{array}\right)=
U\left(\begin{array}{c} \ket{1}\\ \ket{2}\end{array}\right),\quad U=\left(\begin{array}{cc}
\cos\theta & \sin\theta e^{i\phi}\\
-\sin\theta &  \cos\mathcal{\theta} e^{i\phi}
\end{array}\right).
\end{equation}

	At this point, it is necessary to emphasize the importance of the phase $\phi$. The neutrino is the only massive and electrically neutral elementary fermion. Its nature is therefore ambiguous: neutrino can be Dirac or Majorana (\cite{natureNeutrino},\cite{Petcov}). In the Dirac case, the phase $\phi$ is not physical and can be easily removed by rephasing  the neutrino fields occurring in the weak  lepton charged current. On the other hand, in the Majorana case $\phi$ remains and is considered as real and measurable quantity. 
 
Let us now consider Hamiltonian $H$ describing neutrino propagating in matter, which is divided into kinetic ($H_{0}$) and interaction  ($H_{int}$) parts:
\begin{equation}
\label{ham}
H=H_{0}+H_{int}.
\end{equation}
In the ultrarelativistic limit energy of an individual neutrino $E_{i}$ can be expressed as (\cite{Giunti}):
\begin{equation}
E_{i}=E+\frac{1}{2}\frac{m_{i}^{2}}{E^{2}},
\end{equation}
where $E$ is an average energy of a neutrino. Then $H_{0}$ can be, in the mass basis, expressed as function of $\Delta m_{21}^{2}=m_{2}^{2}-m_{1}^{2}$ \cite{Smirnov}:
\begin{equation}\label{h0}
H_{0}=\left(\begin{array}{cc}E+\frac{1}{2}\frac{\Delta m_{21}^{2}}{4E} &0\\
0 & E-\frac{1}{2}\frac{\Delta m_{21}^{2}}{4E} \end{array}\right).
\end{equation}
The potential part arises due to the coherent forward scattering of the electron neutrinos with the electrons (charged current) and all kinds of neutrinos with the electrons, neutrons and protons (neutral current).
Since the contribution from the neutral current is the same for all flavours, the real influence on the neutrino oscillation comes only from the charged current $V_{CC}$:
\begin{eqnarray}\label{hint}
H_{int}=U^{\dag}\left(\begin{array}{cc} V_{CC} & 0\\ 0 &0  \end{array}\right)U=\\
=\frac{V_{CC}}{2}\left(\begin{array}{cc} 1+\cos2\theta & e^{-i\phi}\sin 2\theta\\\nonumber
e^{i\phi}\sin2\theta & 1-\cos2\theta \end{array}\right),
\end{eqnarray}
where $U$ is the mixing matrix given in Eq.(\ref{Eq1}).

It can be easily seen that Hamiltonian given in Eq.(\ref{ham}) does not allow for any discrimination between Dirac and Majorana neutrino: its transformation to the flavour basis makes the phase $\phi$  disappear. However, the treatment of the neutrinos propagating in matter as an open quantum system can dramatically change the situation.

In general, the Markovian evolution of a system interacting with an external environment can be described by the Lindblad-Kossakowski master equation, which in the Heisenberg picture reads as (\cite{kossakowski,Lindblad,breu,alicki}):
\begin{eqnarray}
\label{lindblad}
\frac{d A(t)}{dt}=L^{*}[A(t)],\quad
L^{*}[A(t)]=i[H,A]+\\\nonumber +\frac{1}{2}\sum_{i,j=0}^{n^{2}-1} c_{ij} (F_l^{\dag}[A,F_{k}]+[F_{l}^{\dag},A]F_{k}),
\end{eqnarray}
where $A$ stand for an observable and  $F=\{F_{0},\;F_{1},\;,\cdots, F_{n^{2}-1}\}$ denotes a set of the matrices creating a basis in the $n$-dimensional space.

Note that the right-hand side of Eq.(\ref{lindblad}) is split into two parts. 
In the first part one recognizes the ordinary Heisenberg equation, whereas the second part is responsible for additional effects connected with an interaction of a system  with an environment. As for latter part, in order to assure  the trace- and hermicity-preserving evolution, the coefficients $c_{ij}$ must obey the following conditions \cite{Romano,breu}:
\begin{equation}\label{kond}
|c_{ij}|\leq \frac{1}{2} (c_{ii}+c_{jj}),\quad i,j=1,\cdots n^{2}-1.
\end{equation}

Neutrino during its propagation interacts weakly with matter. It is therefore
subjected to the decoherence and dissipation effects, which justifies the application of Eq.(\ref{lindblad}). Since our system is two-dimensional, the master equation is presented as ($\hbar=1$):
\begin{equation}
\label{heis}
L^{*}(A)=i[H,A]+\sum_{i,j=0}^{3}c_{ij} (\sigma_{j}^{\dag} A \sigma_{i}-\frac{1}{2} \{\sigma_{j}^{\dag},\sigma_{i}\}A),
\end{equation}
where $\sigma_{i}$ denotes one of the set of matrices ($ \mathbb{I}_{2}, \sigma_{x}$,$\sigma_{y}$,$\sigma_{z}$) for $i=0,1,2,3$, respectively.

Assuming the Hamiltonian given in Eq.(\ref{ham}) one can easily simplify the form of Eq.(\ref{heis}) by rearranging it to the Schr\"odinger-like form:
\begin{equation}
\label{alform}
\frac{dA(t)}{dt}=(\mathcal{H}+\mathcal{D})A(t)=\mathcal{H}_{eff}A(t),
\end{equation}
where $A(t)=\sum_{i=0}^{3} a_{i}\sigma_{i}=\left(\begin{array}{c}a_{0}(t)\\a_{1}(t)\\a_{2}(t)\\a_{3}(t)\end{array}\right)$ is a vector represented in the basis of the sigma matrices. Let us notice that Eq.(\ref{alform}) is nothing more than a convenient way of rewritting the Lindblad Master equation Eq.(\ref{lindblad}). 

As for the matrix $\mathcal{H}_{eff}$, it consists of two parts. The dissipative part $\mathcal{D}$ can be parametrized with six real independent parameters \cite{Benatti}:
\begin{equation}
\mathcal{D}=-2\left(\begin{array}{cccc}
0 & 0 & 0 & 0\\
0 & a & b & c \\
0 & b & \alpha & \beta\\
0 & c & \beta & \gamma    \end{array}\right),
\end{equation}
where elements of $\mathcal{D}$ has got the following correspondence with the $[c_{ij}]$ matrix elements:
\begin{gather}
\label{hamiltonnew}
\mathcal{D}_{11}=c_{22}+c_{33},\;
\mathcal{D}_{22}=c_{11}+c_{33},\\\nonumber
\mathcal{D}_{33}=c_{11}+c_{22},\;
\mathcal{D}_{12}=\mathcal{D}_{21}=-c_{12},\\\nonumber
\mathcal{D}_{13}=\mathcal{D}_{31}=-c_{13},\;
\mathcal{D}_{23}=\mathcal{D}_{32}=-c_{23}.
\end{gather}
Subsequently, the direct calculation of the $\mathcal{H}$ matrix elements reveals that:
\begin{gather}
\label{hamiltonnew}
\mathcal{H}_{12}=-\mathcal{H}_{21}=-\frac{\Delta m_{21}^{2}}{2E}+ V_{CC}\cdot \cos2\theta,\\\nonumber
\mathcal{H}_{13}=-\mathcal{H}_{31}=-V_{CC}\cdot \sin\phi \sin2\theta,\\\nonumber
\mathcal{H}_{23}=-\mathcal{H}_{32}=V_{CC}\cdot \cos \phi \sin2\theta.
\end{gather}

The remaining elements in the $\mathcal{H}$ and $\mathcal{D}$ matrices are all equal to $0$. 

It is now easy to solve Eq.\eqref{alform}, for which the formal solution has an exponential form:
\begin{equation}
\label{solution}
A(t)=e^{\mathcal{H}_{eff}t}A(0).
\end{equation}

Let us notice that in principle it is possible to derive the  coefficients $\mathcal{D}_{ij}$  starting from microscopic description of a system including its environment~\cite{alicki}. However, for the systems, which requires description based on the relativistic field theory  this task is often very challenging~\cite{calz}.  

\section{Leggett--Garg  $K_3$ function for the neutrino}

The Leggett-Garg inequalities are usually constructed for the dichotomic (i.e. whose eigenvalues are either +1 or -1) observable $\hat{Q}$ for which quantity defined as:
\begin{equation}
\label{correlation}
C_{ij}=<\hat{Q}(t_{i})\hat{Q}(t_{j})>
\end{equation}
constitutes a benchmark of the correlations between its measurements performed in the different times $t_{i} < t_{j}$ \cite{review}.
Then, the Leggett-Garg parameter $K_{3}$:
\begin{equation}
\label{K3}
K_{3}=C_{21}+C_{32}-C_{13},
\end{equation}
defined for 3 times $t_{3} > t_{2} > t_{1}$ is lower and upper-bounded under the MR and NIM assumptions \cite{review}:
\begin{equation}\label{lgi}
-3\leq K_{3}\leq 1,
\end{equation}
which is known as the simplest from the family of the Leggett-Garg inequalities.

It is shown that for $2$--level system (the kind of which is considered in this paper) the correlation function $C_{ij}$ can be expressed as the expectation value of the symmetrised product \cite{fryc,review,emary}:
\begin{equation}
\label{correlation2}
C_{ij}=\frac{1}{2}\bra{\phi}\{{\hat{Q}(t_{i}),\hat{Q}(t_{j}) }\}\ket{\phi},
\end{equation}
where $\{\hat{A},\hat{B}\}$ states for the anticommutator of the operators $\hat{A}$ and $\hat{B}$.
Then, representing an observable $\hat{Q}$ in the basis of sigma matrices as $\hat{Q}(t_{i})=\vec{q}(t_{i})\cdot\vec{\sigma}$,
 $\vec{\sigma}=(\sigma_{0}=\mathbb{I}_{2},\;\sigma_{1},\sigma_{2},\sigma_{3})$ one arrives to the simplified formula for the coefficient $C_{ij}$ \cite{kaiser}:
\begin{equation}
C_{ij}=\vec{q}(t_{i})\cdot \vec{q}(t_{j}).
\end{equation}
It is important to point out that in the above-introduced formulas we made an assumption of the \emph{stationarity} of the system (no explicit time dependence in the Lindblad equation governing time evolution of the neutrino system Eq.(\ref{alform})), according to which  the values of the correlation functions $C_{ij}$ are not the functions of the specific times $t_{i}$ and $t_{j}$ but rather the time interval $\tau=t_{i}-t_{j}$. Such a requirement accounts to the formal fulfillment of the NIM condition \cite{huelga}.

In order to observe the behaviour of the LGI for the neutrinos propagating in the interactive environment, it is convenient to choose $Q(0)=\sigma_{z}$ as a dichotomic observable. Note that, since
\begin{equation}
Q\ket{\nu_{e}}=\ket{\nu_{e}},\quad Q\ket{\nu_{\mu}}=-\ket{\nu_{\mu}}.
\end{equation}
it measures the neutrino flavour as projected on the z-axis \cite{kaiser}.

Making use of the general solution (Eq.(\ref{solution})) to the Lindblad-Kossakowski master equation (Eq.(\ref{heis})) for  $A(t)=Q(t)$ we are able to examine the variability of  the parameter $K_{3}$ (Eq.\eqref{K3}) as a function of different quantities. In particular, we  focus on a difference 
\begin{equation}\label{DK}
\Delta K_3 = K_3(\phi=0)-K_3(\phi)
\end{equation}
between the $K_3$ function calculated for the Dirac neutrino, corresponding to vanishing CP--violating phase $\phi=0$, and the Majorana with $\phi\neq 0$. Non--vanishing $\Delta K_3$ is a potential quantifier, a hallmark,  indicating different behaviour of these two types of neutrinos. 

\begin{figure}
  \includegraphics[width=1.0\linewidth]{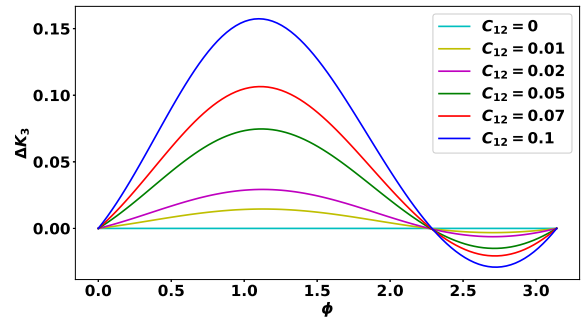}
\caption{The difference $\Delta K_{3}$ as a function of time for Dirac and Majorana neutrinos versus $\phi$ for different values of $c_{12}=c_{21}$, fixed potential $V_{CC}=2$ and $c_{11}=c_{22}=c_{33}=0.1$ with $E=1$.}
\label{fig1}
\end{figure}

An influence  of decoherence for the properties of $\Delta K_3$  for equal time intervals $\tau$   and for different values of the off--diagonal elements $c_{12}=c_{21}$, with  remaining parameters (such as $E$ and $V_{CC}$) fixed, is presented in Fig. (\ref{fig1}).  As we set $\hbar=1$ all other parameters are expressed in suitable powers of $eV$. In particular, in our calculations we set $\tau=0.1$ and assume the experimental values of the solar mixing angle $\theta_{12}=0.187\pi$ and $\Delta m_{21}^{2}\approx 7.54\cdot 10^{-5}$ \cite{olive}.  Furthermore, the values of the $[c_{ij}]$ Kossakowski matrix, rather than derived from any fundamental properties of the system, such as the coupling with the well-described thermal environment \cite{lobejko,alicki},  result from a phenomenological modeling \cite{Benatti,Dajka}. Our choice of the non--vanishing $c_{ij}$, yet the simplest possible,   has no qualitative influence on the presented results. Increasing either the number or the value (provided that Eq.(\ref{kond}) is satisfied) of non--vanishing matrix elements $c_{ij}$ simply lowers the amplitude of $\Delta K_3$. 

Let us explicitly state here the central result of our work:   non--vanishing of any of the off--diagonal elements of the Kossakowski matrix $c_{ij}\neq 0$  is {\it necessary} for $\Delta K_3\neq 0$. In other words, any '{\it off}--diagonal decoherence' is a {\it sine qua non} condition for $\Delta K_3$ to be able to   discriminate Dirac and Majorana nature of oscillating neutrinos. 

\begin{figure}
  \includegraphics[width=1.0\linewidth]{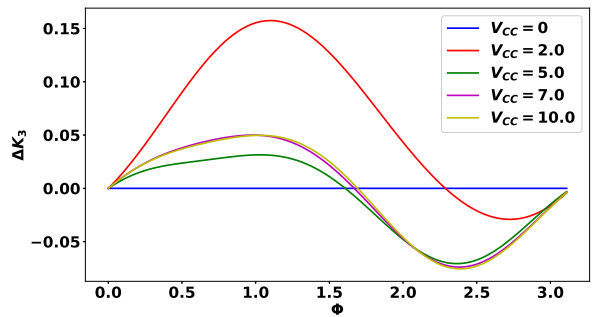}
  \caption{The difference $\Delta K_{3}$ as a function of time for Dirac and Majorana neutrinos for different values of potential $V_{CC}$, given energy $E=1$ and elements of the Kossakowski matrix $c_{12}=c_{21}=c_{11}=c_{22}=c_{33}=0.1$  with $E=1$.}
\label{fig2}
 \end{figure}

In Fig.(\ref{fig2}) we present   the parameter $\Delta K_{3}$ as function of $\phi$ influenced by different values of $V_{CC}$ quantifying interaction of neutrinos with the normal matter. The  results of the numerical tests for Dirac  and Majorana  neutrinos bring us to a conclusion that there is an optimal value of $V_{CC}$ (in the case of  parameters chosen here $V_{CC}\approx 2$) for which the difference $\Delta K_3$ is maximal and  allows for the most efficient discrimination between Dirac and Majorana neutrino's nature. For our parameters the difference is the most apparent for $V_{CC}\approx 2$, whereas is  hardly visible for $V_{CC}<0.6$ and $V_{CC}>7.0$.  As  the potential $V_{CC}$ describes interaction of neutrinos with matter one can conjecture that this type of interaction and, in particular, a presence of matter, can enhance the possible implications of the neutrino's nature at least on certain measurable properties. 

\begin{figure}
  \includegraphics[width=1.0\linewidth]{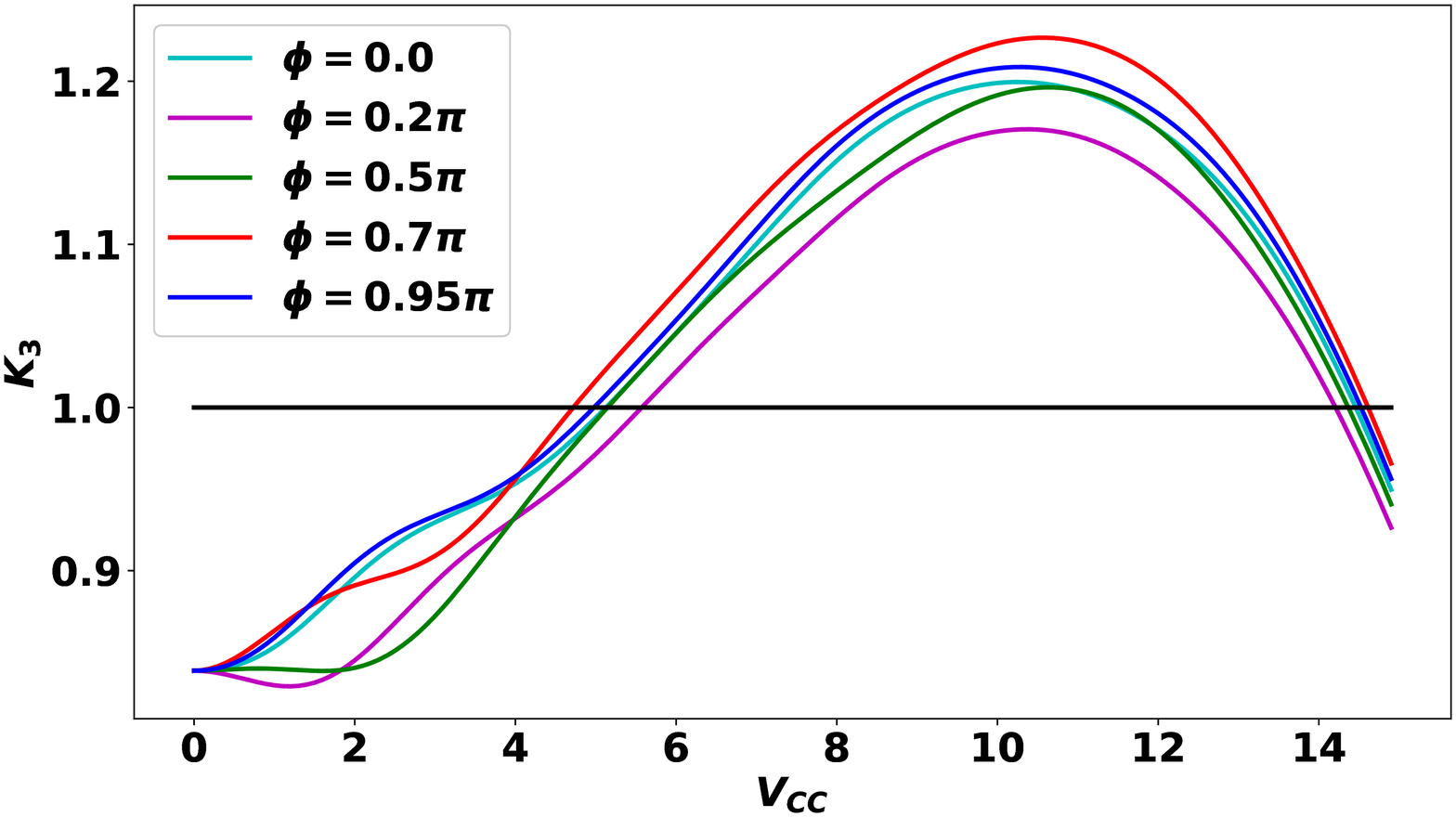}
  \caption{The dependence of the coefficient $K_{3}$ on the potential $V_{CC}$  in the dissipative ($c_{12}=c_{21}=c_{11}=c_{22}=c_{33}=0.01$) environment  with $E=1$. }
\label{fig3}
 \end{figure}

In studies of the $K_3$ correlator, it is natural to investigate possibility of violation of the LGI Eq.(\ref{lgi}) in dissipative neutrino systems. We investigate whether the condition $K_3>1$ may be satisfied more likely either for Dirac or for Majorana neutrinos. The first observation is that for both  non-dissipative environments (with $c_{ij}=0$) and the diagonal  Kossakowski matrix,   there is no possibility to distinct from Dirac and Majorana neutrinos solely upon studies of LGIs' violation.
  On the other hand, from  Fig.(\ref{fig3}) one infers that for {\it off}--diagonal decoherence 
there exists  a certain value of the  CP-violating phase $\phi$ for which the value of $K_3$ is maximized. Similarly, there is also an optimal value of $V_{CC}$, in our case $V_{CC}\approx 10$, such that the violation of the LGI is most significant. 
Let us notice, however, that the  value of $V_{CC}$ which leads to a maximal violation of the LGI is different comparing to that which results in a maximal value of $\Delta K_3$ discussed previously. According to results of our numerical analysis not reproduced here this property is  valid universally i.e. for any choice of parameters of the system.   

\begin{figure}
  \includegraphics[width=1.0\linewidth]{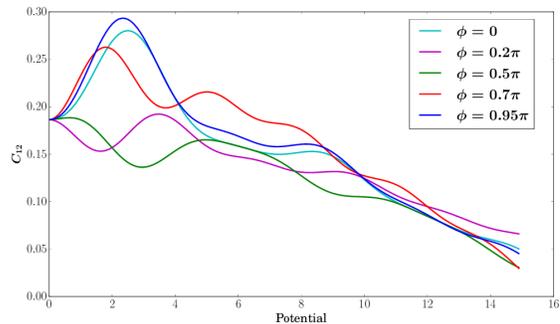}
  \caption{The dependence of the correlation function  $C_{ij}$ Eq.(\ref{correlation2}) on the potential $V_{CC}$  in the dissipative ($c_{12}=c_{21}=c_{11}=c_{22}=c_{33}=0.01$) environment  with $E=1$. }
\label{fig4}
 \end{figure}

\begin{figure}
  \includegraphics[width=1.0\linewidth]{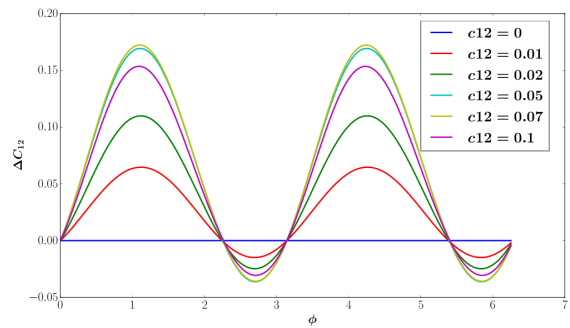}
  \caption{The difference of the correlation function  $\Delta C_{ij}=C_{ij}(\phi=0)-C_{ij}(\phi)$ for Dirac and Majorana neutrinos  for different values of potential $V_{CC}$, given energy $E=1$ and elements of the Kossakowski matrix $c_{12}=c_{21}=c_{11}=c_{22}=c_{33}=0.1$  with $E=1$.}
\label{fig5}
 \end{figure}

Let us notice that both the $K_3$ LG--correlator Eq.(\ref{K3}) and the difference $\Delta K_3$ in Eq.(\ref{DK}) are formed by a very special combination of correlation functions $C_{ij}$ defined in  Eq. (\ref{correlation2}).  It is worth to mention that 
all the properties of the LG--correlators which seem to be useful for distinguishing Dirac and Majorana neutrinos essentially originate form the properties of $C_{ij}$ and $\Delta C_{ij}=C_{ij}(\phi=0)-C_{ij}(\phi)$ as presented in Fig. (\ref{fig4}) and Fig. (\ref{fig5}) respectively.  However, the additional physical content present in the LG--correlator $K_3$ supported by recent experimental efforts on studies of a LGI violation in neutrino systems~\cite{kaiser} is encouraging to  relate this phenomenon -- quantified by $K_3$ rather than $C_{ij}$ -- and the investigation of the neutrino's nature.

Eventually, let us notice that, contrary to $V_{CC}$,  an  energy $E$ in Eq.(\ref{h0}) -- one of these parameters which are crucial for the properties of oscillating neutrinos --  affects neither $\Delta K_3$ nor $K_3$. The chosen value of $E$ in  Fig. (\ref{fig1}), Fig. (\ref{fig2}) and Fig.(\ref{fig3}) is completely arbitrary.  In other words, one cannot either make Dirac and Majorana neutrinos distinguishable or improve their distinguishability solely via setting a value of $E$ in a similar manner as it can be done  with $V_{CC}$. 

\section{Conclusions}
Determination of the neutrino nature, if it is Dirac or Majorana particle,  is one of the most important, still unsolved, problems in studies of these fascinating particles. There are various problems arising in this context. First of all, the measurement of any of the neutrinos' properties is a hard task due to a well known weakness of the neutrino's interactions. Secondly, one needs to find and measure an observable, which would allow for the establishment of the neutrino nature. In this theoretical  work we attempt to contribute to this second aspect. We show that the $K_3$ correlator studied recently in a context of violation of the Leggett--Garg inequalities takes different values for Dirac and Majorana neutrinos under certain conditions which were identified in our work. We studied neutrino's oscillation in a presence of decoherence and matter using phenomenological approach based on the Kossakowski--Lindblad--Gorini completely positive dynamics~\cite{alicki}. 

The main conclusion of our modelling is that the difference in $K_3$ occurs 
 provided that {\it (i)} there is a certain type of decoherence represented by an {\it off}--diagonal Kossakowski matrix  and {\it (ii)} the neutrino interacts with a matter i.e.  $V_{CC}\neq 0$ in Eq.(\ref{hint}). The two conditions are necessary for making the difference $\Delta K_3$ non--vanishing and potentially useful candidate for a neutrino's nature quantifier.


We also studied the difference in a  possible violation of the LGI for Dirac and Majorana 
neutrinos. We concluded that the same two conditions as formulated above are valid also for the $K_3>1$ Leggett--Garg condition. However, there is a crucial difference: the value of $V_{CC}$ necessary for violation of the LGI ($K_3>1$) is one order of magnitude larger then the corresponding value necessary for $\Delta K_3\neq 0$. As the value of $V_{CC}$ is related to density of matter affecting neutrino's oscillations, this observation may be potentially interesting and useful for experimentalists. 

Let us emphasize here that credibility of any  quantitative results obtained via any phenomenological model are questionable as long as the model is not confronted with a huge amount of experimental data. As neutrino physics suffers from a shortage of experimental data {\it quantitative} predictions presented in our work require considerable caution in interpretation.  One can not exclude a very pessimistic scenario such that the proposed quantifier discriminating neutrinos nature  works in a tailored range of parameters which is essentially inaccessible in any direct experiment and that our predictions are of purely theoretical character. 

Mysterious neutrinos, more than any other 'fundamental' particles, attract physicists outside the Particle Physics Community. We hope that our results are not only a modest contribution to physics of neutrinos {\it per se} but also indicate some unusual properties of neutrinos  shedding light on even more unusual properties of the Quantum World.


\section*{Acknowledgments}
The  work has been  supported   by the NCN project
UMO-2013/09/B/ST2/03382 (B.D. and M.R.) and the
NCN grant 2015/19/B/ST2/02856 (J.D) .

\bibliography{biblio}

\end{document}